# Three-body Simulations of a Primordial Black Hole Encounter with the TOI-2796 System

S.V. Vereshchagin[1, *], M. I. Kamentsev[2]

[1]*Institute of Astronomy Russian Academy of Sciences (INASAN), 48 Pyatnitskaya st., Moscow, Russia*
[2]*Omsk State Technical University, 11 Mira st., Omsk, Russia*
*E-mail: svv@inasan.ru

**Abstract.** We present numerical three body simulations of a primordial black hole (PBH) encounter with the TOI 2796 system, which hosts a hot Jupiter. Over 100 days, we identify three outcomes: (i) ejection of the planet; (ii) a bound triple system (a approx. 1 au, P approx. 5 days); and (iii) capture of the planet by the PBH into a binary (0.0194 au, 15 days) that escapes the star. These results highlight the rich dynamical diversity of PBH encounters and their potential role in shaping planetary systems.

**Key words:** exoplanet, planet system, TOI-2796, Primary Black Hole, three-body problem

## INTRODUCTION

A primordial black hole (PBH) is a hypothetical type of black hole that formed not from the gravitational collapse of a star (as is the case for conventional black holes) but from overdense regions of matter during the earliest stages of the Universe's evolution. They are thought to have emerged within the first fractions after the Big Bang, arising from fluctuations in the gravitational field or matter density. PBHs are considered promising candidates for the dark matter (DM) component of the Universe (Villanueva-Domingo et al., 2021).

The mass of PBH can span an exceptionally broad range—from microscopic scales to supermassive values. They can be searched for via gravitational waves from mergers (Escrivà et al., 2024), gravitational microlensing (the bending of light from distant stars), and the products of their Hawking evaporation (Auffinger, 2023). Additional signatures of DM activity in the Solar system have been proposed. De Rocco (2025) notes that DM in the form of macroscopic composites remains largely unconstrained in the mass range $10^{11} - 10^{17}$ g. Within this mass window, DM may collide with planetary bodies, depositing energy and leaving striking surface features that remain detectable over geological timescales. He suggests that Ganymede, Jupiter's largest moon, serves as an excellent target for detecting such collisions, owing to its differentiated internal structure and a surface aged over billions of years. Scholtz and Unwin (2020) proposed that anomalies in the orbits of trans-Neptunian objects (sometimes attributed to a hypothetical Planet 9) could be explained by a PBH that was either captured by or recently passed through the Solar system. Brown et al. (2025) investigated the influence of primordial black holes on exoplanetary systems.

If PBHs constitute a fraction of DM, a natural question arises: how frequently do PBHs enter the Solar system? For a Solar system bounded by Neptune's orbit, the answer is rarely. However, for a system extended to the size of the Oort cloud, the answer differs. Simulations of DM bombardment of the Oort cloud (Mould, 2026) have demonstrated that the ejection of comets from the Oort cloud and their injection into the inner Solar system could match the observed cometary frequency, provided the PBH fraction is sufficiently high.

Tran et al. (2024) examined how the gravity of a passing object perturbs a planet's orbit and which characteristic signatures could, in principle, be extracted from precise orbital data. According to simulations using the REBOUND code (Brown et al. 2025), even rare PBH flybys can leave a detectable "gravitational fingerprint" in the form of orbital anomalies, with lower-mass planets being particularly vulnerable. The possibility of PBH capture (alongside free-floating planets) by a protostellar cloud was considered by Eroshenko (2023).

In this work, we examine the dynamical response of the TOI-2796 planetary system to a close flyby of a hypothetical primordial black hole (PBH) by means of direct three-body numerical integrations. Our simulations, carried out over a 100-day interval, reveal that the planet TOI-2796b can undergo either complete ejection, capture into a bound PBH–planet pair that escapes the star, or incorporation into a stable triple system composed of the PBH, the star, and the planet. These findings illustrate the diverse gravitational outcomes of PBH–planetary encounters and underscore the potential of PBHs to act as agents of planetary system disruption.

## PHYSICAL PARAMETERS OF THE OBJECTS UNDER STUDY

### *Star TOI-2796*

Star TOI-2796 (Keele University, URL=https://www.astro.keele.ac.uk/jkt/tepcat/planets/TOI-2796.html, Southworth 2026). TOI-2796 (also known as TIC 220076110) is a single star of spectral type G2–G3 V, located at a distance of about 352.4 pc from the Sun. Equatorial coordinates RA=$05^h36^m36.65^s$, DEC=+00°53′46.6″, proper motion 3.73 ± 0.10 mas/yr, mass 1.063 (+0.076, -0.062) $M_\odot$, radius 1.069 ± 0.024 $R_\odot$, metallicity [Fe/H] ≈+0.239 (+0.076, -0.079), apparent magnitude (V)~12.36–12.53 mag, age~4.0 Gyr.

### *The Exoplanet TOI-2796 b*

The Exoplanet TOI-2796 b (Keele University, URL=https://www.astro.keele.ac.uk/jkt/tepcat/planets/TOI-2796.html, Southworth 2026). TOI-2796 b is a gas giant (a "hot Jupiter") discovered in 2022–2023 from TESS data. Parameters: mass 0.44 (+0.10, -0.11) $M_{Jup}$ (≈ 139 $M_\oplus$), radius1.59 (+0.05, -0.50) $R_{Jup}$ (≈ 17.82 $R_\oplus$), orbital period 4.8084983 ± 0.0000057 days, semi-major axis 0.0569 (+0.0010, -0.0011) au, eccentricity 0.0 (+0.172, -0.0), orbital inclination~84.9°. The planet orbits its star closer than Mercury orbits to the Sun.

### *The Hypothetical Primordial Black Hole*

Unlike the well-constrained host star and its planet, the primordial black hole (PBH) is a purely hypothetical object, and its properties are not fixed by observations. In our dynamical study, we therefore treat the PBH parameters as free variables within physically motivated ranges. In general, a PBH characterized by its mass, which can span an extraordinarily broad spectrum—from asteroid-scale masses, as low as $\sim 10^{-8} M_\odot$ , up to several hundred solar masses, depending on the formation epoch.

The mass of a PBH is intimately linked to the cosmic time at which it formed: during the first fractions of a second after the Big Bang, the horizon mass sets the typical PBH mass. Because lighter black holes evaporate faster via Hawking radiation, only those with initial masses exceeding $\sim 10^{15}\ g \approx 5 \cdot 10^{-19} M_\odot$ could have survived to the present day; all lighter PBHs would have completely evaporated by now. As reviewed by Carr et al. (2019), the most promising mass windows for PBHs that remain viable today are $10^{16} - 10^{17} g, 10^{20} - 10^{24} g, 10 - 10^3 M_\odot$. These ranges are not exclusive but reflect theoretical and observational constraints. The physical size of a PBH is determined by its Schwarzschild radius, Rs=$2GM/c^2$. For a PBH of one solar mass, Rs≈3km; for lower masses, the radius scales proportionally. In our simulations, the PBH is treated as a standard point mass, with its gravitational field fully described by Newtonian gravity (we neglect relativistic corrections, as the typical impact distances are much larger than the Schwarzschild radius).

The PBH's initial velocity and impact parameter relative to TOI-2796 are treated as free parameters in our encounter model. We explore a representative range of these parameters to evaluate both the immediate orbital perturbations of the planet and the possible destabilization of a putative comet cloud around the star. This latter effect is particularly motivated by the system's close passage near the NGC 1977 cluster about 5.5 Myr ago (Vereshchagin and Chupina, 2024), which may have already shaped the outer regions of the system.

## A COMPUTATIONAL FRAMEWORK

The numerical integration was performed using the *odeint* routine from *SciPy*. This routine implements an adaptive integrator based on the LSODA (Livermore Solver for Ordinary Differential Equations with Automatic Method Switching for Stiff and Non-Stiff Problems) algorithm (Hindmarsh and Petzold 2005). In practice, *scipy.integrate.odeint* is a wrapper around the Fortran LSODA library. LSODA automatically switches between the Adams method (suitable for non-stiff problems, such as the three-body problem under normal conditions) and the BDF (Backward Differentiation Formula) methods, which are employed if the system becomes stiff.

### *Equations of Motion for the Three-Body Problem*

We solve the Newtonian equations of motion for three gravitating point masses. For each body *i* with mass $m_i$, position vector $\boldsymbol{r}_i$, and velocity $\boldsymbol{v}_i = \dot{\boldsymbol{r}}_i$ the acceleration is given by

$$\ddot{\boldsymbol{r}}_i = \sum_{i \neq j} \frac{G m_i (\boldsymbol{r}_j - \boldsymbol{r}_i)}{|\boldsymbol{r}_j - \boldsymbol{r}_i|^3},$$

where $G$ is the gravitational constant. In our code, we work in astronomical units (au), solar masses $M_\odot$, and days. The gravitational constant was adopted as G = 0.00029598739 in these units ($\mathrm{au}^3 M_\odot^{-1} \mathrm{day}^{-2}$), (G = $4.3\times10^{-3}$ pc$\times M_\odot^{-1}$(km/s)$^2$), all velocity components must be expressed in units of au $\mathrm{day}^{-1}$. To use the *scipy.integrate.odeint* routine, we rewrite this second-order system as a first-order ordinary differential equation (ODE) system. We define a state vector **y** of length 18 for three bodies:

$$\mathbf{y} = [x_1, y_1, z_1, x_2, y_2, z_2, x_3, y_3, z_3, v_{x1}, v_{y1}, v_{z1}, v_{x2}, v_{y2}, v_{z2}, v_{x3}, v_{y3}, v_{z3}],$$

where $x_i, y_i, z_i$ are the coordinates of body *i*, and $v_{xi}, v_{yi}, v_{zi}$, where the last three terms are the velocity components. The time derivative of this vector is

$$\dot{\mathbf{y}} = \begin{bmatrix} v_{x1}, v_{y1}, v_{z1}, v_{x2}, v_{y2}, v_{z2}, v_{x3}, v_{y3}, v_{z3} \\ a_{x1}, a_{y1}, a_{z1}, a_{x2}, a_{y2}, a_{z2}, a_{x3}, a_{y3}, a_{z3} \end{bmatrix},$$

where the accelerations $a_{xi}, a_{yi}, a_{zi}$ are computed from the gravitational force sums given above.

### *Error Control*

For the three-body problem, which is inherently chaotic, error control is considerably more important than the specific choice of the Adams method. The accuracy is governed by the tolerances *rtol* (relative) and *atol* (absolute). Explicitly specifying these parameters can substantially improve the stability of the integration. In our case rtol=1e-8, atol=1e-8.

It is worth noting that under our environment (Windows 7), no manual step-size control is required; the step size is chosen automatically, even during close encounters between bodies, while the built-in error control continuously monitors the numerical accuracy. For exploratory computations in *SciPy*, this approach is a standard and reliable choice.

TIME EVOLUTION OF THE TOI-2796 SYSTEM UNDER PBH FLYBY

The visualization module displays three-dimensional trajectories and their projections onto the xy-plane. The code is entirely self-contained and does not require special astronomical packages. It allows easy modification of masses, initial positions and velocities, and enables investigation of various dynamical regimes of the three-body problem.

The computed trajectories of the three interacting bodies—the host star, the planet TOI-2796b, and the PBH—are shown in Figures 1–3. Each figure corresponds to a different set of initial conditions, specifically varying the PBH impact parameter and mass. These plots reveal the rich diversity of post-encounter configurations, including planetary ejection, capture by the PBH, and formation of a bound triple system.

*Ejection of the Planet from the System*

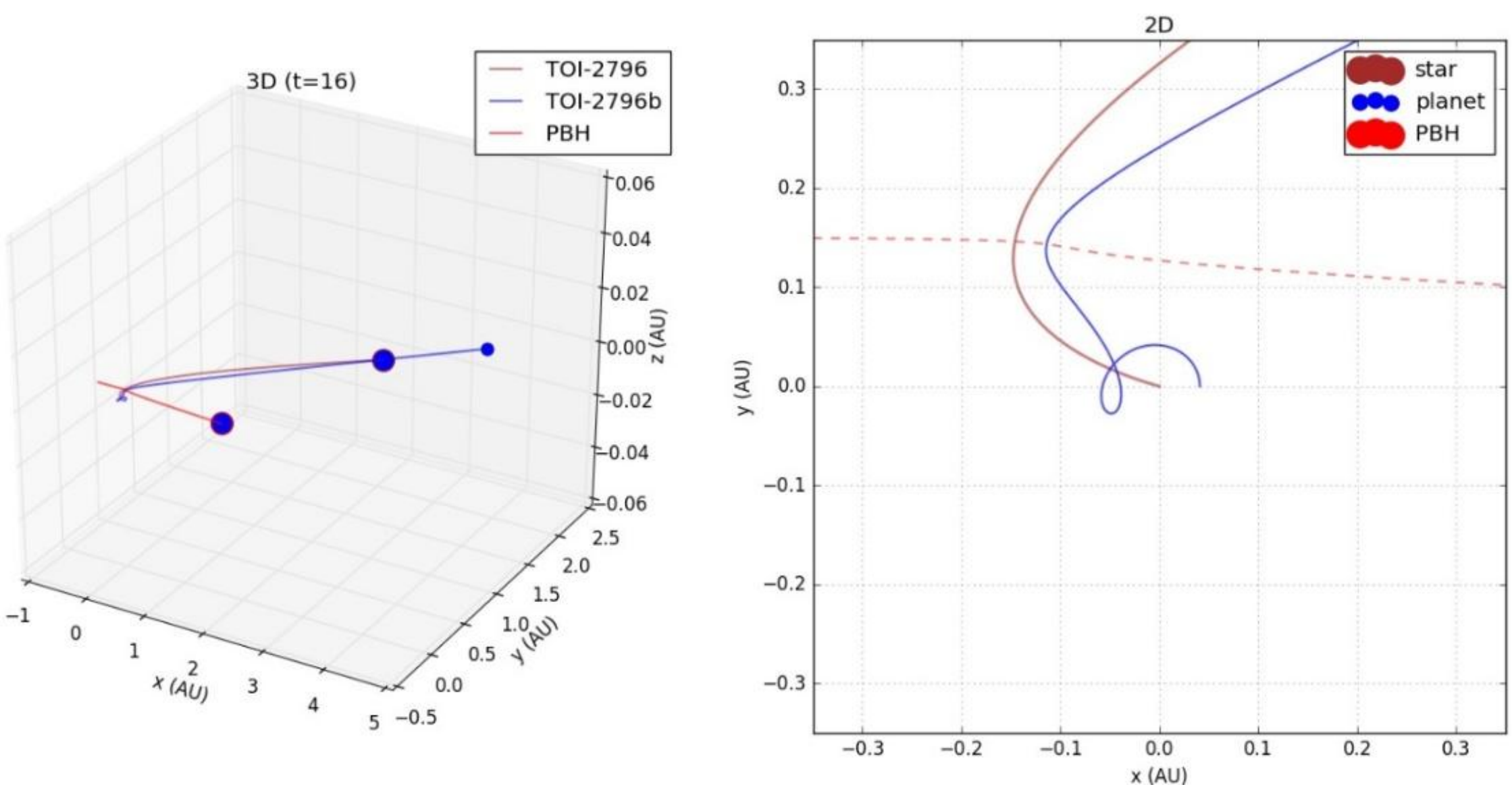

Fig. 1. Close encounter leading to planetary ejection. The left panel shows the three-dimensional motion of the three bodies. Following the PBH flyby, the star loses its planet, which recedes to a distance of about 2 au over the 16-day integration interval. The right panel presents a zoomed-in 2D view of the region around the closest approach, where the strongly diverging paths of the unbound objects are clearly visible.

On Fig.1 the PBH mass is $15.0 M_{\odot}$, and its initial position is (x,y,z)= [-0.6, 0.15, 0.0] au, the number of integration steps was 5000.

*Formation of a Three-Body Gravitational System - Evolution toward a Bound Triple System*

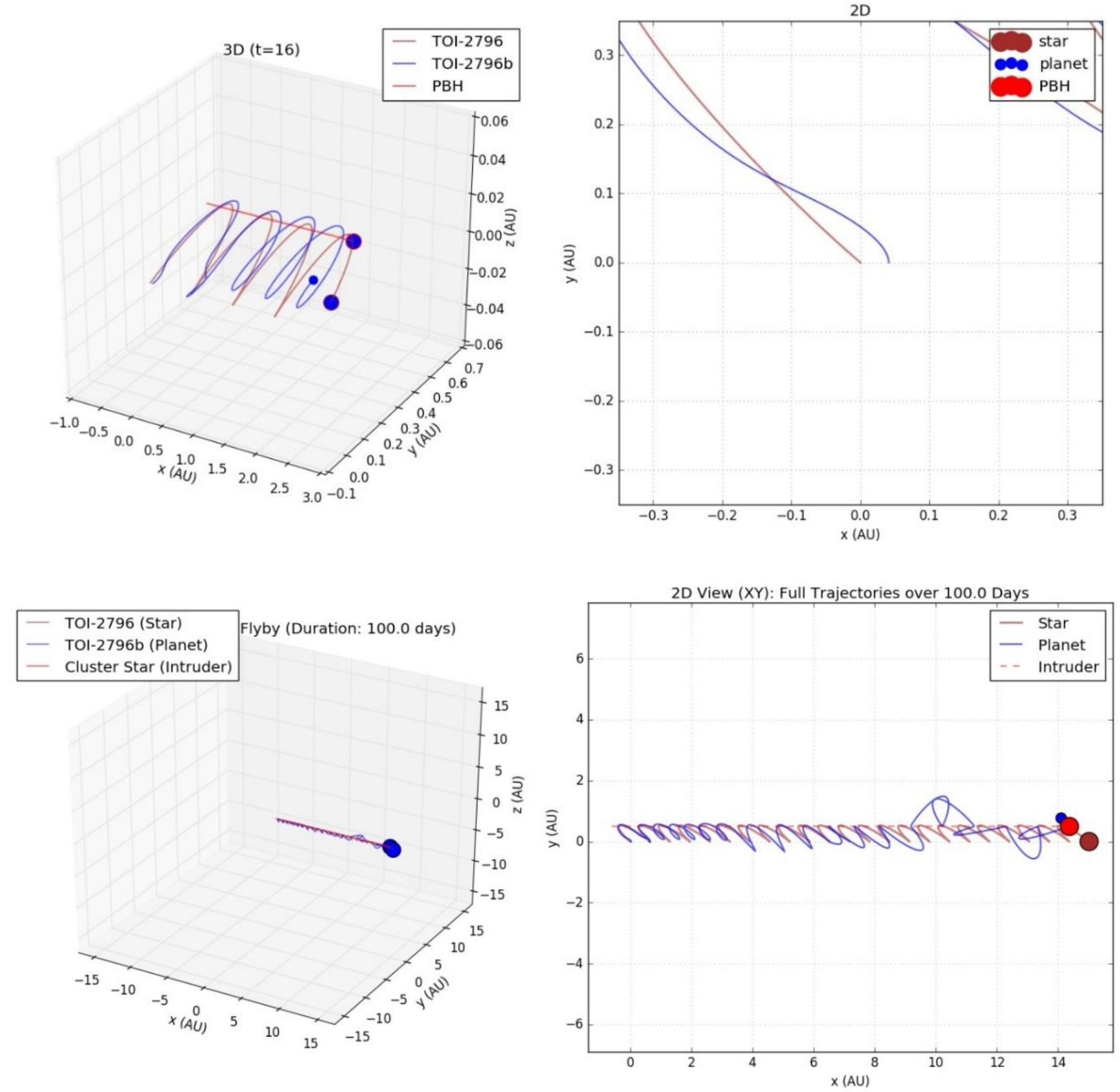

**Fig. 2.** Close encounter leading to the formation of a bound triple system — evolution toward a stable hierarchical configuration. The left panels show three-dimensional trajectories of the three bodies; the right panels present the corresponding projections onto the xy-plane. The PBH mass is 500.0 $M_{\odot}$, and its initial position is (x,y,z)= [-0.6, 0.5, 0.0] (in au). The upper panels correspond to an integration time of 16 days, illustrating the immediate post-encounter dynamics, while the lower panels show the evolution over 100 days, demonstrating the long-term stability of the resulting triple system. The estimated PBH–star orbital elements are a≈1 au and P≈5 days; the motion shows a gradual increase in a and noticeable perturbations.

From the trajectories on Fig. 2, we estimate the semi-major axis of the PBH–star orbit to be a≈1 au, with an orbital period of about 5 days. The plots also reveal a gradual increase in the semi-major axis and noticeable irregularities in the motion, indicative of the ongoing gravitational interactions within the triple system.

### *Formation of a PBH–Planet Bound Pair*

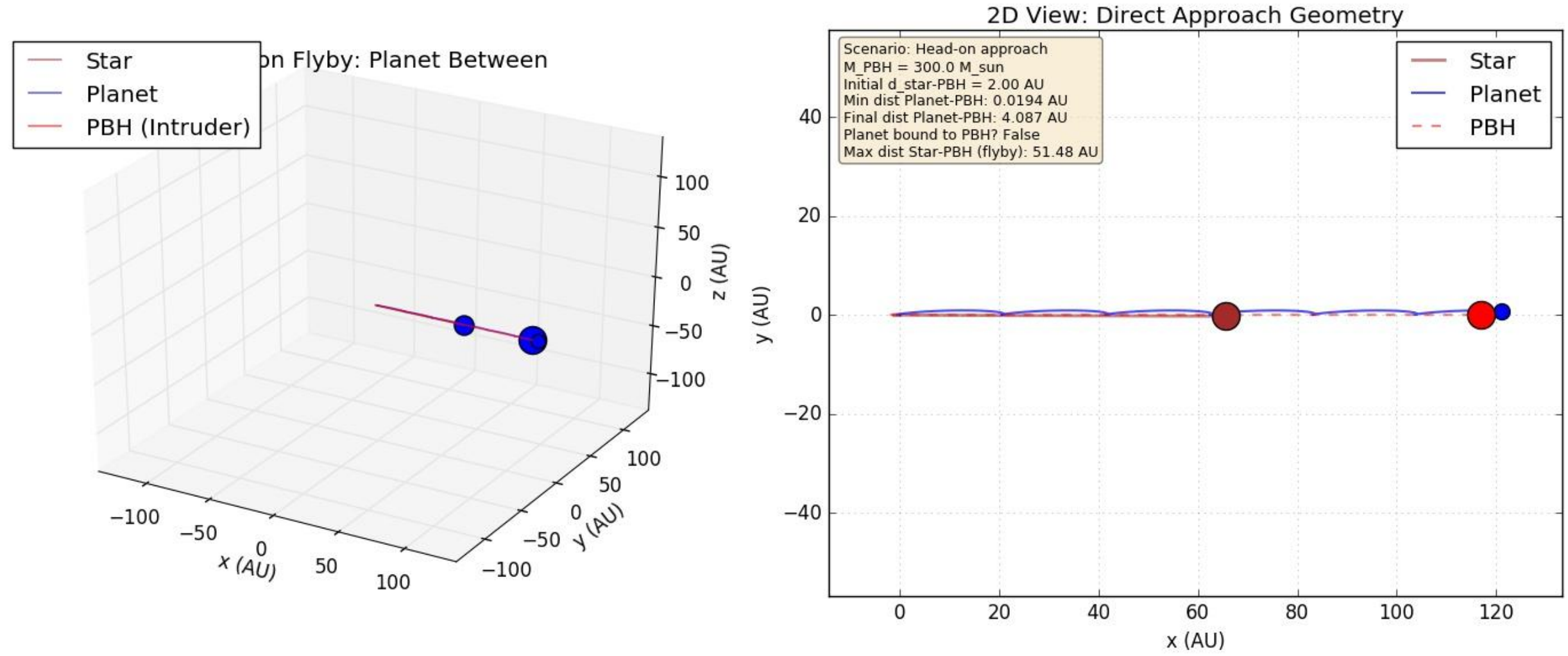

Fig. 3. Capture scenario the formation of a PBH–planet bound pair. The left panels show three-dimensional trajectories of the three bodies; the right panels present the corresponding xy-plane projections. The PBH mass is 300.0 $M_{\odot}$, and its initial position is (x,y,z)-in text (in au). The integration time is 100 days. After the encounter, the planet becomes gravitationally bound to the PBH, forming a compact binary with a semi-major axis of approximately 0.0194 au and an orbital period of about 15 days. Over the integration interval, the PBH–planet pair recedes to a distance of about 51.5 au from the host star, effectively escaping the original system.

To obtain the result shown in Fig. 3 we trained a neural network to search for a specific post-encounter configuration following the PBH flyby. A total of 475 simulation runs were carri ed out, and the network successfully identified the captured PBH–planet binary presented in Fig. 3.

Parameters obtained from the neural network training:

```
Computing scenario: PBH flies directly towards Star, Planet is in between.
Scenario: Head-on approach
M_PBH = 300.0 M_sun
Initial d_star-PBH = 2.00 AU
Min dist Planet-PBH: 0.0194 AU
Final dist Planet-PBH: 4.087 AU
Planet bound to PBH? False
Max dist Star-PBH (flyby): 51.48 AU
```

## DISCUSSION AND CONCLUSIONS

### *Single-Planet Systems*

Systems hosting a single planet are not rare but rather appear to be the norm. According to NASA Exoplanet Archive data, more than 2,000 known planetary systems currently have only one confirmed planet. However, it is not the systems themselves that attract the greatest interest, but rather their unusual properties and the puzzles surrounding their origin. The key question is: why does such a system contain only one planet? The single-planet status may be temporary; additional planets could be too small or too distant to be detected with current observational techniques. Nevertheless, studies indicate that a substantial fraction of hot Earths are indeed found in single-planet systems. This suggests that there exist unknown physical processes that either inhibit the formation of additional planets in these systems or destroy them at early evolutionary stages.

Notable examples of planetary systems with only one detected planet include the following. The nearest single-planet system, Gliese 832, is a red dwarf with a single confirmed planet. Another system, MOA-2010-BLG-477L, a Jupiter analogue with a mass of 1.4±0.3 $M_{Jup}$ orbiting at a semi-major axis of 2.8±0.5 au, reminiscent of the orbit of our own gas giant.

*Observational Searches for Primordial Black Holes*

Strictly speaking, no such objects have been detected to date. Possible signatures of primordial black holes (PBHs) include orbital anomalies, microlensing, and accretion flares. However, each of these signals either remains unconfirmed upon repeated scrutiny or admits a more mundane explanation.

Both the Vera C. Rubin Observatory and the Nancy Grace Roman Space Telescope are poised to revolutionize the search for primordial black holes (PBHs), primarily through the detection of gravitational microlensing—the transient brightening of a background star caused by the passage of a massive compact object (URL=https://www.nasa.gov/missions/roman-space-telescope/how-nasas-roman-mission-will-hunt-for-primordial-black-holes/)

The "Phoebe" case. A notable episode occurred in 2023, when researchers reported a transient brightening of a star in the Large Magellanic Cloud, which was interpreted as gravitational microlensing by a PBH of roughly lunar mass. Subsequent re-analysis of the data, including long-term OGLE monitoring (Hawkins and García-Bellido 2025), revealed that the star is intrinsically variable (Udalski, Mróz 2026), and its natural flickering produces such spikes (Niikura, H., et al. 2019). A genuine microlensing event from a single passage does not repeat, whereas the observed recurrent brightness variations provided the key to resolving the nature of the signal.

Orbital anomalies. Researchers have modelled how a PBH flyby could distort exoplanetary orbits. The concept is physically sound: the gravity of a close massive object does indeed alter orbital parameters such as eccentricity and inclination (Brown et al. 2025). In practice, however, current telescopes do not measure the orbits of most exoplanets with sufficient precision to reliably detect the imprint of a single brief event against the background of other perturbations (e.g., from neighbouring planets or tidal forces).

Accretion flares. If a PBH traverses a gas cloud or stellar wind, an accretion disk may form and produce a flare (Kalita, Maity 2025). Yet such events are difficult to attribute unambiguously to PBHs, as similar signals can arise from other compact objects or variable stars.

*Concluding Remarks and Summary of the Identified Systems*

The dynamical evolution was investigated within the framework of the three-body problem. Using numerical integrations within the framework of the restricted three-body problem, we have investigated the dynamical consequences of a primordial black hole (PBH) passing in the vicinity of the TOI-2796 planetary system. The TOI-2796 system, consisting of a Sun-like star and a single hot Jupiter, was chosen as a representative test case. Our numerical integrations within the three-body framework have shown that a close passage of a PBH through the TOI-2796 system can lead to qualitatively different dynamical outcomes on a timescale of 100 days. By systematically varying the initial conditions, we obtained the following outcomes:

Ejection of the planet from the system. The planet became unbound from the host star on a timescale of approximately ten days, reaching a distance exceeding 2 au from the star.

Formation of a bound triple system. A stable triple configuration emerged, with a characteristic size of about 1.0 au and an orbital period of roughly 5 days, as derived from integrations covering 100 days.

Capture of the planet by the PBH. To identify specific post-encounter configurations, we trained a neural network on 475 simulation runs. The network successfully identified a scenario in which the planet became gravitationally bound to the PBH, forming a PBH–planet binary. The resulting orbital separation was 0.0194 au, with an orbital period of about 15 days. Over the integration time, the PBH–planet pair receded to a distance of 51.48 au from the host star.

For more complex cases, it's worth trying to adopt the *solve ivp* interface, which allows explicit selection of integration methods such as RK45 (Runge–Kutta 5(4)) or DOP853 (an explicit Runge–Kutta method of order 8). In particular, the DOP853 method provides high accuracy, essential for long-term integrations with minimal energy drift.

## ACKNOWLEDGEMENTS

This work made use of the LSODA solver from the ODEPACK library, developed by Hindmarsh and Petzold (2005). The solver was called through the *scipy.integrate.odeint* interface. LSODA automatically switches between non-stiff (Adams) and stiff (BDF) methods, providing reliable integration for the three-body problem. The authors thank John Southworth (Keele University, UK) for making the TEPCat data (Southworth 2026) on the TOI-2796 system publicly available. The authors thank T. B. Keskin for helpful assistance.